\documentclass[spanish]{maciarticle}

\usepackage{amsmath,amsbsy,amscd,amssymb,graphicx,epsfig,color}

\begin{document}

\title{Réplica del valor de un pool (CPM) y hedging de pérdidas impermanentes}

\author[$\dag$]{A. Muñoz González}
\author[$\dag$,$\ddag$]{J.I. Sequeira}
\author[ ]{Ariel Dembling}
\affil[$\dag$]{Facultad de Ciencias Exactas y Naturales,  Universidad de Buenos Aires, Intendente Güiraldes 2160, Ciudad Universitaria, Núñez, Argentina, \textcolor{blue}{https://exactas.uba.ar/}}
\affil[$\ddag$]{IMAS-CONICET, Godoy Cruz 2290, Ciudad Autónoma de Buenos Aires, Argentina, \textcolor{blue}{https://www.conicet.gov.ar/}}
%

 \maketitle

\begin{abstract}
Este artículo caracteriza analíticamente la pérdida impermanente para los creadores automáticos de mercado en mercados descentralizados como Uniswap o Balancer (CPMM). Presentamos una fórmula teórica de replicación estática para el valor del pool  mediante una combinación de calls y puts europeos. Formularemos un resultado para garantizar cobertura para todo precio final que se encuentre en un intervalo predefinido. 
\end{abstract}
\begin{mathsubclass}
91G15
\end{mathsubclass}

{\thispagestyle{empty}} 

\section{Introducción} \label{s:intro}


Un market maker simple pero sorprendentemente eficaz parece ser el creador de mercado de producto constante (CPMM) utilizado por Uniswap \cite{zhang2018formal} y Balancer. Los creadores de mercado automáticos (AMM) se han estudiado ampliamente en la teoría de juego algorítmica, comenzando con la regla de puntuación de mercado logarítmica de Hanson (LMSR) \cite{hanson2003combinatorial}. Dichos AMM se construyen primero haciendo que los proveedores de liquidez depositen activos en una proporción fija para especificar una distribución inicial de creencias sobre los posibles resultados. Luego, un AMM proporciona una regla de puntuación que especifica el costo de cambiar la distribución de creencias de su estado actual a un nuevo estado deseado. 

El proveedor de liquidez está expuesto a una pérdida impermanente que solo se realiza hasta que se agota la liquidez y se retiran los tokens del pool.  Dado que los traders siempre intercambian tokens menos valiosos por uno más valioso, los proveedores de liquidez  sufren pérdidas impermanentes (IL) que podrían ser significativas.

En este trabajo, proponemos una estrategia de cobertura estática para proveedores de liquidez utilizando opciones europeas estándar para eliminar el impacto de la IL. En primer lugar, mostramos que los proveedores de liquidez tienen una exposición que es equivalente a mantener posiciones largas y cortas en diferentes opciones de calls y puts. 
Luego mostramos un resultado que asegura la protección de la impermanent loss en un intervalo $[P_i, P_s]$ siempre que ciertas desigualdades sobre las cantidades y costos de puts y calls a comprar, el capital total invertido y los retornos obtenidos del pool, sean satisfechas.

\section{Motivacion: Réplica estática de un payoff con opciones y bonos}
Para la siguiente sección usamos \cite{carr2001towards} que revisa la teoría de replicación estática usando opciones desarrolladas por Ross \cite{ross1976options} y Breeden y Litzenberger \cite{breeden1978prices}.


  Cualquier función $f:\mathbb{R}_+\to \mathbb{R}$ que mapea el precio $p$ de algún activo riesgoso a una cantidad $f(p)$, si pedimos que sea dos veces diferenciable se escribe 
 
 \begin{equation}\label{replicador con opciones}
     f(P_T)=f(m)+f'(m)[(P_T-m)^+-(m-P_T)^+]+\\
     \int_{0}^{m}f''(K)(K-P_T)^+dK +\int_{m}^{\infty} f''(K)(P_T-K)^+dK,
 \end{equation}
donde $m\in\mathbb{R}^+$. El primer término se  interpreta como el pago de una posición estática en $f(m)$ bonos de descuento puro (exentos de riesgo de impagos), cada uno de los cuales paga un dólar en $T$. El segundo término se puede interpretar como el payoff de $f'(m)$ calls con strike $m$ menos el payoff de $f'(m)$ puts, también con strike $m$. El tercer término como una posición estática en $f'' (K)dK$ puts en todos los strikes menores a $m$. De manera similar, el cuarto término como una posición estática en $f'' (K)dK$ calls en todos los strikes mayores que $m$.

En ausencia de arbitraje, una descomposición similar a \eqref{replicador con opciones} se debe mantener entre los valores iniciales. Sean $V_0 ^{f}$ y $B_0$ los valores iniciales del payoff y del bono de descuento puro, respectivamente; $P_0(K)$ y $C_0(K)$ los precios iniciales de la opción put y call en $K$, respectivamente. Entonces el valor presente de un payoff como el anterior es 

 \begin{equation}\label{replicador con opciones inicila time}
     V_0 ^f=f(m)B_0+f'(m)[C_0(m)-P_0(m)]+\\
     \int_{0}^{m} f''(K)P_0(K) dK+\int_{m}^{\infty} f''(K)C_0(K)dK.
 \end{equation}
Por tanto, el valor de un payoff arbitrario se puede obtener a partir de los precios de bonos y opciones.






\section{Hedging contra las pérdidas impermanentes}

\subsection*{Definicion}
Para una provisión de liquidez con depósitos $x$ e $y$ de los tokens $X$ y $Y$ en el momento inicial $0$, la pérdida impermanente realizada (IL) al retirar la liquidez en el momento $t$, es la pérdida de capital en comparación con haber mantenido el par de tokens estáticamente en el momento inicial $0$. Específicamente, la pérdida impermante (IL) se calcula como

$$IL = V_{pool} - V_{hold} = (y_t + x_t*P_t) - (y_0 +x_0 * P_t)$$
donde, $x_t$ e $y_t$ son las cantidades retiradas en el tiempo $t$ y $P_t$
es el precio de una unidad del token $X$ en unidades de $Y$.

\subsection*{Fórmulas para IL}
Si consideramos un pool de liquidez (CPM), se tiene
$$xy=k, \ \frac{y}{x}=P.$$
A partir de estas ecuaciones se deducen las siguientes 
$$x=\sqrt{\frac{k}{P}}, \ y=\sqrt{k P},$$
y ahora  calculemos IL en función de precios
\begin{align*}
    V_{Pool}(P) := 
    V_{pool}(P_0)\sqrt{\frac{P}{P_0}}, \
    V_{Hold}(P) := 
    \frac{V_{Hold}(P_0)}{2}\left(\frac{P}{P_0}+1\right)
\end{align*}
donde  $V_{Pool}(P_0)=V_{Hold}(P_0):=y_0+x_0 P_0$, tenemos 
$$IL(P) = V_{Pool}(P)-V_{Hold}(P)=
V_{Hold}(P_0)\left(\sqrt{\frac{P}{P_0}}-\frac{1}{2}\left(\frac{P}{P_0}+1\right)\right).$$
 
 \begin{figure}[ht]\label{IL}
    \centering
    \includegraphics[width=15cm]{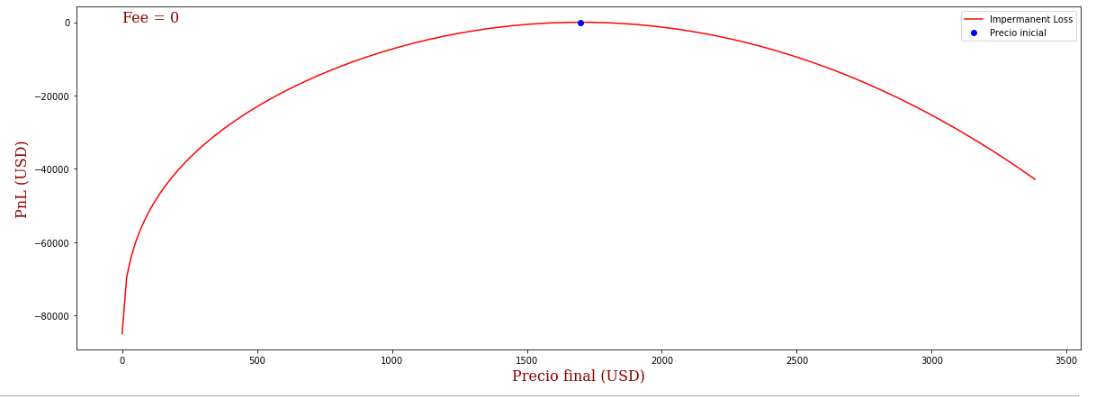}
    \caption{Pérdidas impermanentes}
\end{figure}
    
Podemos entonces calcular la derivada de $IL(P)$ con respecto a $P$:
\begin{align*}
    \frac{\partial}{\partial P} IL(P) = V_{Hold}(P_0)\left(\frac{\partial}{\partial P}\left(\sqrt{\frac{P}{P_0}}\right)-\frac{1}{2P_0}\right)
     = \frac{V_{Hold}(P_0)}{2P_0}\left(\sqrt{\frac{P_0}{P}}-1\right)
\end{align*}

Por lo tanto, para cada tiempo $t$
$$IL(P_t)=V_{Hold}(P_0)\left(\sqrt{\frac{P_t}{P_0}}-\frac{1}{2}\left(\frac{P_t}{P_0}+1\right)\right),\ 
\frac{\partial}{\partial P} IL(P_t)=\frac{V_{Hold}(P_0)}{2P_0}\left(\sqrt{\frac{P_0}{P_t}}-1\right).$$

En particular, notar que $IL(P_0)=\frac{\partial}{\partial P} IL(P_0)=0$.
\section{Resultados} \label{s:methods.3}
\subsection*{Hedging estático de IL con una estrategia del tipo Long Strangle }
Un proveedor de liquidez ingresa al pool de producto constante depositando $x_0$  unidades de USDC e $y_0$ unidades de Ethereum (ETH). Sea  $r_p$ la tasa de retorno mensual que paga el pool. A tiempo $T$ (un mes por ejemplo) tenemos cierta pérdida impermanente por depositar los tokens en el pool
$$IL(P_T)= V_{pool} - V_{hold} = c \left(\sqrt{\frac{P}{P_0}}-\frac{1}{2}\left(\frac{P}{P_0}+1\right)\right),$$
donde $c=:  x_0+y_0 P_0$ es el capital inicial en dólares depositado en el pool y $P_0$ es el precio en dólares de ETH en el momento que se ingreso.

Si consideramos $f=IL$, $m=P_0$, aplicamos el resultado de replicación estática y observamos que $IL(P_0)=IL'(P_0)=0$, entonces la ecuación \eqref{replicador con opciones} nos dice que para replicar la IL debemos comprar puts y calls con strikes a izquierda y derecha del precio inicial $P_0$ resp. Debido a la falta de liquidez de derivados financieros en el ecosistema cripto, en la práctica resultaría costoso (o directamente imposible) realizar compras continuas como aparece en la ecuación. 
Más aún, para tener una disponibilidad más alta de derivados en el mercado y reducir los costos al máximo lo que haremos será replicar la IL sólo en un intervalo de movimiento del precio considerando un strike para los puts y un strike para los calls. 

La estrategia consistirá entonces en comprar $q_c$ unidades de opciones calls con strike $K_c$  y $q_p$ unidades de puts con strike $K_p$, con primas $d_c$ y $d_p$ resp.

El PnL de la estrategia con opciones a tiempo $T$ es  $$q_c(P_T-K_c)^+ +q_p(K_p-P_T)^+ - D$$  
donde $D:=q_c d_c+q_p d_p$ es el costo total de las opciones.

Por lo tanto, el PnL del depósito en el pool junto con la cobertura a tiempo $T$ es

$$ f_T^{\textit{pool+str}}= r_p c+\textit{payoff}_{str}-D +IL(P_T).$$
Para no estar en pérdida buscamos que sea mayor o igual que cero.
 
Debido a que compraremos puts y calls en un sólo strike, no está claro que la cantidad óptima a comprar sea $IL''(K)$. La siguiente propiedad define condiciones para estas cantidades:

\begin{figure}\label{Hedge}
    \centering
    \includegraphics[width=12cm]{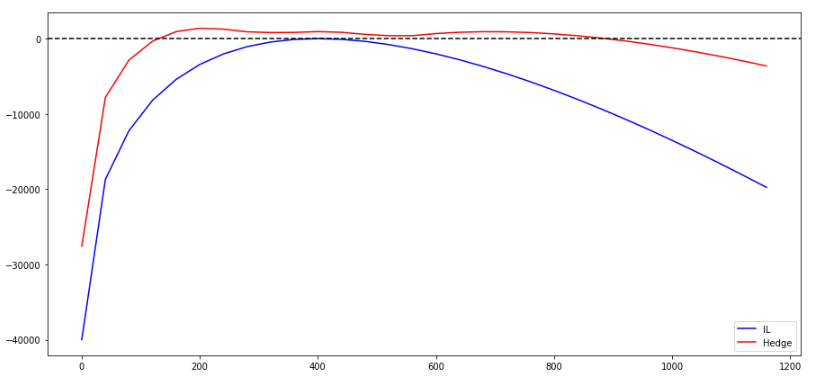}
    \caption{PnL Estrategia vs Movimiento del precio}
\end{figure}

\subsubsection*{Proposición}
 Sean $P_0$ precio de Ethereum a tiempo $0$, $c$ capital en dólares depositado en el pool a tiempo $0$ y $r_p$ radio de retorno que paga el pool a tiempo $T$.
Para cubrir (pnl no negativo) las pérdidas impermanentes a tiempo $T$ en un intervalo $[P_i, P_s]\subset \mathbb{R}_{\geq 0}$ (que contiene a $P_0$) con una estrategia del tipo Long Strangle (con opciones europeas) y posiciones Long en calls (si $q_c>q_p$) ó puts (si $q_p>q_c$), basta que se cumplan simultáneamente las siguientes desigualdades:
\begin{align*}
    \frac{c}{2}\left(\frac{1}{\sqrt{P_iP_0}} - \frac{1}{P_0}\right)&\leq q_p\\
     D-\min\{IL(K_c),IL(K_p)\}&\leq r_p c\\
    -\frac{c}{2}\left(\frac{1}{\sqrt{P_s P_0}}-\frac{1}{P_0}\right)&\leq q_c
\end{align*}
donde $K_c$,$K_p$ son strikes,  $d_c$,$d_p$ primas y  $q_c$, $q_p$ cantidades correspondiente a opciones europeas  call y puts respectivamente con fecha de expiración $T$, por último  $D=q_c d_c+ q_p d_p$ corresponde al costo de la estrategia, Ver Fig. \ref{Hedge}

\section{Conclusión}\label{s:conclusion}
En este trabajo, hemos mostrado que si uno conoce el retorno que el pool ofrece y dada una cantidad $c$ de capital a invertir, es posible definir un rango de movimiento del precio en el cual estaremos protegidos, y estimar la cantidad mínima de posiciones en opciones put y calls (con strikes igual a los bordes del rango de hedgeo, respectivamente) que debemos tomar. 


Finalmente, notar que la réplica estática de la sección 2 es útil en la teoría pero no tanto en la práctica debido a los costos y la falta de oferta de derivados financieros en el mercado de las criptomonedas. Si bien el resultado de la sección anterior asegura una protección en un cierto rango, la aplicabilidad de dicha estrategia dependerá nuevamente de que dichos instrumentos estén disponibles para los retailers.

\end{document}